\documentclass[aps,prstab,superscriptaddress,floatfix,twocolumn,amsfonts,amsmath,amssymb,amsart]{revtex4-1}

\usepackage{comment}
\usepackage{tikz-feynman}
\usepackage{graphicx} 
\usepackage{float}
\usepackage{dcolumn} 
\usepackage{bm} 
\usepackage[colorlinks, breaklinks=true,linkcolor=red, citecolor=blue, linktocpage=true]{hyperref}

\begin{document}
\renewcommand{\vec}{\mathbf}
\renewcommand{\Re}{\mathop{\mathrm{Re}}\nolimits}
\renewcommand{\Im}{\mathop{\mathrm{Im}}\nolimits}

\title{Excitons in Atomically Thin TMD in Electric and Magnetic Fields}
\author{Jack N. Engdahl}
\thanks{j.engdahl@student.unsw.edu.au}
\affiliation{School of Physics, University of New South Wales, Sydney 2052, Australia}
\author{Harley D. Scammell}
\affiliation{School of Mathematical and Physical Sciences, University of Technology Sydney, Ultimo, NSW 2007, Australia}
\author{Dmitry K. Efimkin}
\affiliation{School of Physics and Astronomy, Monash University, Victoria 3800, Australia}
\author{Oleg P. Sushkov}
\affiliation{School of Physics, University of New South Wales, Sydney 2052, Australia}

\date{\today}

\begin{abstract}
The magnetic field dependence of photoabsorption provides direct insights into the band structure of semiconductors. It is perhaps surprising that there is a large discrepancy between electron, hole, and reduced mass reported in the recent literature. Motivated by this puzzle we reconsider excitonic magneto-absorption and find that the commonly employed perturbative approach, namely for computing the diamagnetic shift, is inadequate to account for the parameter ranges considered in existing data.  In particular, we develop the theory for strong magnetic field and, upon analysis of the data, arrive at the set of exciton parameters different to what has been estimated perturbatively in the literature.  Only s-wave excitons are visible in  photoluminescence as the spectral weight of p-wave states is too small, this limits the amount of information that can be extracted about the underlying band structure.  To overcome this, we propose to study p-wave states by mixing them with s-wave states by external in-plane electric field and show that a moderate DC electric field would provide sufficient mixing to brighten p-wave states. We calculate energies of the p-wave states including the effects of valley-orbital splitting and the orbital Zeeman shift, and show that this provides direct information on the electron-hole mass asymmetry.
  \end{abstract}
\maketitle

\section{Introduction}
Monolayer (1L) Transition Metal Dichalcogenides (TMDs) are a class of two dimensional (2D) semiconducting materials that feature a direct band gap at the K points of the Brillouin zone, quadratic band structure and large band gaps~\cite{wang_colloquium_2018,scharf_dynamical_2019}. The strong Coulomb interaction in 1L TMDs results in huge exciton binding energies which causes excitons to dominate the absorption spectrum even at room temperature~\cite{wang_colloquium_2018,scharf_dynamical_2019, chen_luminescent_2019, meckbach_influence_2018,qiu_screening_2016,qiu_giant_2019}. 1L TMD semiconductors are 2D systems with hexagonal Brillouin zones, however they lack inversion symmetry. Both the conduction and valence bands are spin split by the spin-orbit interaction
and the spin and valley degrees of freedom are coupled~\cite{kormanyos_k_2015,stier_exciton_2016,rostami_effective_2013,wang_colloquium_2018,scharf_dynamical_2019}. The unique properties of 1L TMDs render them an ideal platform for the study of exciton physics ~\cite{chen_luminescent_2019,wu_exciton_2015,liu_hybrid_2021,goldstein_ground_2020,liu_landau-quantized_2020,bange_ultrafast_2023,zhu_-plane_2023} and dynamical screening effects, particularly in regards to quasiparticle band gap renormalization~\cite{gao_dynamical_2016,liang_carrier_2015,qiu_giant_2019,qiu_screening_2016,zibouche_gw_2021,liu_direct_2019}. Further, 1L TMDs are ideal for the development of optoelectronic devices~\cite{wang_electronics_2012,mueller_exciton_2018,taffelli_mos2_2021}, including photosensor devices ~\cite{yin_single-layer_2012,velusamy_flexible_2015,chang_monolayer_2014,zhang_highgain_2013,lopez-sanchez_ultrasensitive_2013,gonzalez_marin_mos2_2019} and the fabrication of vertical and lateral heterostructures with advanced optical performance, with such devices providing an excellent environment for light-matter coupling in the form of exciton-polaritons. ~\cite{dufferwiel_valley_2018,zhao_exciton_2023}. As such, accurate determination of the effective Hamiltonian parameters of 1L TMDs would be beneficial for modelling these technologies, and several others, including single photon emitters~\cite{sortino_bright_2021,schuler_electrically_2020,srivastava_optically_2015,tonndorf_single-photon_2015}.

Exciton physics in 2D TMD systems is highly sensitive to reduced (band) mass, characteristic screening length and the external dielectric constant and hence the appropriate analysis of exciton resonances can be used to determine these quantities. Further to this, we will show how the electron and hole band mass asymmetry can also be extracted from excitonic spectroscopy, providing a near complete picture of the underlying band structure (effective Hamiltonian).


In the present work we consider excitons in magnetic and electric fields. While our conclusions are generic,
to be specific we concentrate on WSe$_2$.  There are two major messages
of our work. (i) At realistic out-of plane magnetic fields 
a simple perturbation theory approach in magnetic field is not sufficient and one needs an exact solution to analyse 
existing data. Developing  the solution and performing  analysis of the data we determine parameters of the system which differ significantly from those obtained with simple perturbation theory. (ii) While s-wave TMD excitons have been
observed in a number of photoexcitation experiments, p-wave excitons are
invisible because of the extremely small spectral weight.
Application of a moderate in-plane electric field makes p-wave exciton states visible in  photoexcitation. We predict properties of the p-wave states:
energies, photoexcitation probabilities, valley-orbital level splitting and
orbital Zeeman effect due to the electron-hole mass-asymmetry.
Experimental studies of p-wave states are absolutely feasible and they will allow the study of these effects and to establish rather precisely the mass-asymmetry.

(i)
Measurements of diamagnetic energy shifts of exciton s-states   provides a route to determine the effective parameters of 1L TMDs. 
The perturbation theory formula for the diamagnetic shift reads~\cite{landau_quantum_2007},
\begin{equation}
\label{eq:H_simple}
    \delta E =\frac{e^2}{8 \mu} \langle r^2 \rangle B^2,
\end{equation} 
with $e=|e|$ the elementary charge, $\mu$ the reduced mass, $\langle r^2 \rangle$ the square of the radius of the exciton in a given quantum state, and $B$ the out-of-plane magnetic field strength.
This formula is used to fit experimental data
on TMD excitons~\cite{walck_exciton_1998,chen_luminescent_2019,stier_exciton_2016}. We will show that this formula, obtained from a perturbation theory in $B$, is insufficient at strong $B$-fields. And instead, we will derive the appropriate strong field behaviour of the diamagnetic shift. This was previously shown for a 2D Coulomb problem in magnetic field in the strong field and weak field limits~\cite{macdonald_hydrogenic_1986,laird_rydberg_2022} but in this work we consider a Keldysh potential and present a solution for arbitrary magnetic field, similar to Ref.~\cite{kezerashvili_magnetoexcitons_2021} however we extend the method to account for arbitrary orbital angular momentum and particle-hole mass asymmetry which is essential for real systems.
Using our non-perturbative approach, we can provide accurate fitting to available experimental data, and in doing so, establish the key parameters for the low-energy TMD Hamiltonian.

(ii) Physics of p-wave exciton states is more rich than that of s-wave states.
It includes the valley-orbital effect (the splitting of $l=\pm 1$ states
without magnetic field) and the orbital Zeeman effect (the splitting of
$l=\pm 1$ states in magnetic field). These phenomena are intimately related
to the valley anomalous Hall effect in electron transport.
Unfortunately the p-wave states  are invisible in photoexcitation; these are
``dark'' states.
We propose to make the states visible by applying an in-plane electric field.
The electric field should be sufficuently strong to brighten the p-wave states relative to the s-wave states, but still remain weak enough such that bound states are not destroyed. We find the appropriate electric field strength is $E \sim 1-3$V/$\mu$m.
We calculate values of the aforementioned effects in combined electric and magnetic fields.

The rest of the paper is organised as follows:
Section~\ref{background} discusses the diamagnetic shift and existing
photoluminescence experimental data.
Section~\ref{sec:1} walks through the derivation of the effective Hamiltonian
for exciton in 1L TMD.
In Section~\ref{sec:2} we present our results for s-wave excitons.
We compare our theory to experimental results and extract reduced mass,
dielectric constant and characteristic screening length.
In Section~\ref{sec:EE} we consider excitons in combined in-plane electric and
out-of-plane magnetic fields and show how such measurements can allow the study of
anomalous effects in 1L TMDs.
 We summarise our conclusions in Section~\ref{sec:3}.

\section{Background: The Diamagnetic Shift}\label{background}
\begin{figure}
  \centerline{
  \includegraphics[width=0.46\textwidth]{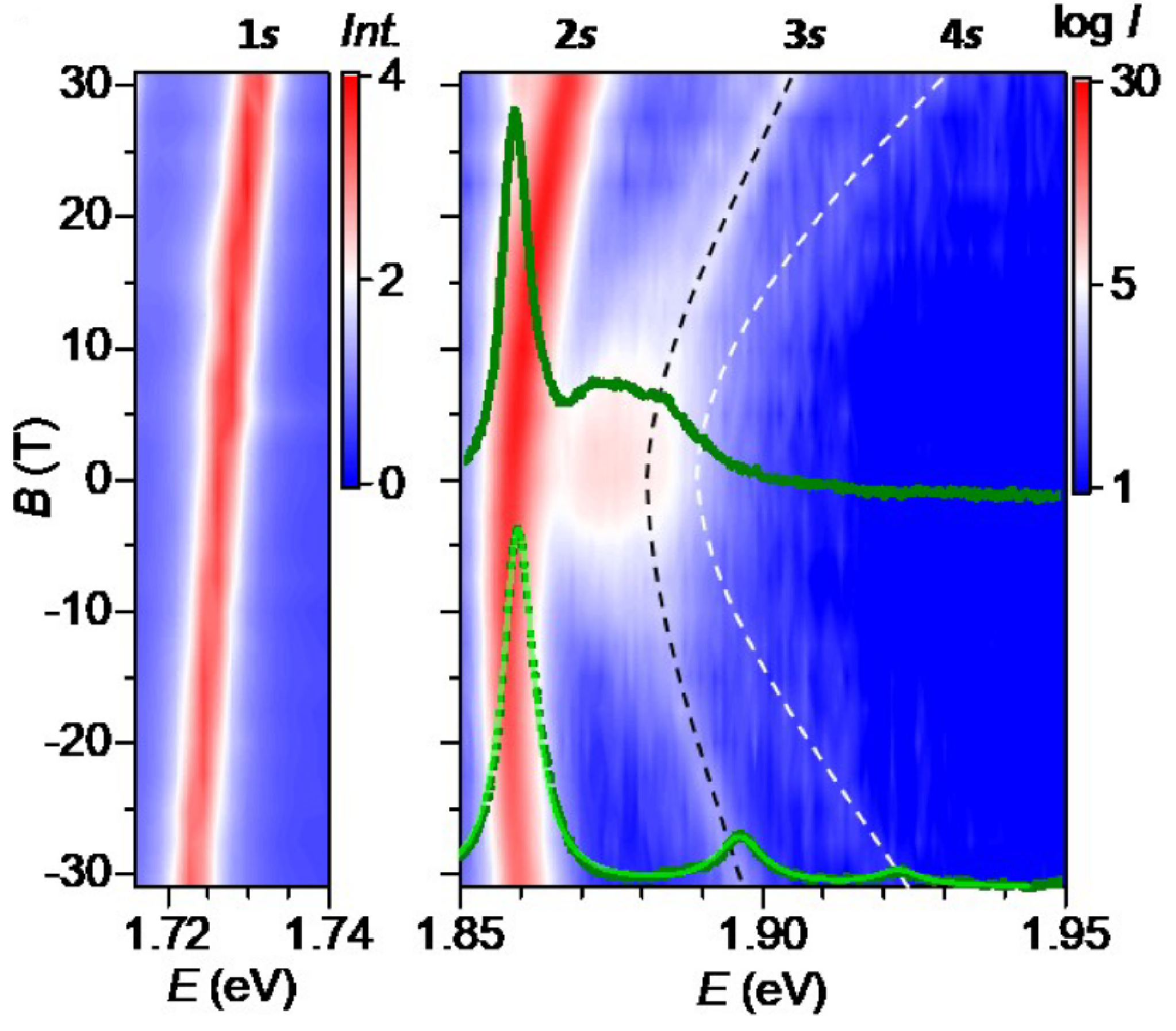}}
  \caption{Exciton energy of s-wave excitons
 in 1L WSe$_2$ encapsulated in  hBN
 in a magnetic field as measured using photoluminescence, taken
 from Ref.~\cite{chen_luminescent_2019}. The colour plot shows the intensity of photoluminescence spectra for 1s, 2s, 3s and 4s excitons.
A particular valley is selected using the circularly polarised light.
 There is a linear valley Zeeman effect that causes an offset between $B<0$ and $B>0$ results. The superimposed green lines are plots of the photoluminescence spectra at $B=0$ T and $B=-31$ T.}	
	\label{fig:exp}
\end{figure}

To stress the shortcomings of Eq.\eqref{eq:H_simple}, we present
Fig.~\ref{fig:exp}, taken from experimental paper Ref.~\cite{chen_luminescent_2019}, which plots the excitonic energies against magnetic field 
in 1L WSe$_2$ encapsulated in hBN.
The data shows a combination of the diamagnetic shift, which is quadratic in B,  and  a
linear in B valley Zeeman effect that we discuss in Appendix \ref{app:zeeman}.
The valley Zeeman effect is not exciton state specific, it is common for all
exciton states. 
To exclude the valley Zeeman effect the data should be symmetrised in B.
It is clear by inspection of Fig.~\ref{fig:exp}
that the 3s and 4s diamagnetic shifts are not quadratic at high $B$ and hence cannot be described by Eq.\eqref{eq:H_simple}; the perturbation theory is insufficient. 
A strong field approach is necessary.

There is an interesting theoretical aspect of the problem that we would like to stress.
In the weak B-field limit the expression for the diamagnetic energy shift,
Eq.\eqref{eq:H_simple}, is independent of the dimensionality of the exciton.
However, in strong $B$-field the exciton dynamics strongly depends on dimensionality.
In the three-dimensional (3D) case the exciton wave function is like a needle
aligned with the magnetic field, and the Larmor circle oscillates along the needle resulting in a series of 1D Coulomb levels built on
each Landau level ~\cite{landau_quantum_2007}.
In the case of a 2D exciton the $z$-confinement does not allow this oscillation. The exciton remains confined to the $x$-$y$ plane at any field~\cite{macdonald_hydrogenic_1986}.
Hence, diamagnetic shifts in strong B are very different in 3D and 2D cases.

\section{Hamiltonian for Exciton in Magnetic Field} \label{sec:1}
In the presence of a magnetic field, the formation of an exciton can be described by the following Hamiltonian 

\begin{equation}
\begin{split}
  \hat{H}=\frac{\left(\hat{\vec{p}}_\mathrm{e}+ \frac{e}{c}{\vec A}_\mathrm{e} \right)^2}{2m_\mathrm{e}} + \frac{\left(\hat{\vec{p}}_\mathrm{h} -\frac{e}{c}
    {\vec A}_\mathrm{h}\right)^2}{2m_\mathrm{h}} + V(\vec{r}_e-\vec{r}_h)
\end{split}
\label{Hamiltonian1}
\end{equation}
Here 
$\hat{\mathbf{p}}_{\mathrm{e}(\mathrm{h})}$ is the momentum operator for the electron (hole), and $m_{\mathrm{e}(\mathrm{h})}$ is its effective mass. In the cylindrical gauge, the magnetic field B can be described by the vector potential $\mathbf{A}_{\mathrm{e}(\mathrm{h})} = [\mathbf{B} \times \mathbf{r}_{\mathrm{e}(\mathrm{h})}]/2$. If TMD monolayer is deposited at the insulating substrate, the Coulomb attraction is accurately described by the Rytova-Keldysh potential given by 
\begin{equation}
V(\vec{r})=-\frac{\pi e^2}{2 \epsilon r_0} \left[H_0\left(\frac{r}{r_0}\right)-Y_0\left(\frac{r}{r_0}\right) \right].
\end{equation}
Here $H_0$ and $Y_0$ are the Struve and Bessel functions of the second kind respectively and $r_0$ is the characteristic screening length. The length $r_0=2\pi \alpha/\epsilon$ is determined by the polarizability of the TMD monolayer $\alpha$ and effective dielectric constant $\epsilon$ of the surrounding media.

It is instructive to introduce the center of mass $\mathbf{R}= (m_\mathrm{e}\mathbf{r}_\mathrm{e}+m_\mathrm{h}\mathbf{r}_\mathrm{h})/(m_\mathrm{e}+m_\mathrm{h})$ and relative $\mathbf{r}=\mathbf{r}_\mathrm{e}-\mathbf{r}_\mathrm{h}$ coordinates.
The corresponding momenta are ${\vec P}=-i\hbar\frac{\partial}{\partial {\vec R}}$ and ${\vec p}=-i\hbar\frac{\partial}{\partial {\vec r}}$.
For the remainder of this work we shall use units $\hbar=c=1$. The Lamb transformation~\cite{lamb_fine_1952} ${\hat H} \to U^{\dag}{\hat H}U$ governed by the unitary operator $U=e^{i \frac{e}{2}\mathbf{B}\cdot[\mathbf{R}\times\mathbf{r}]}$ further simplifies Hamiltonian Eq.(\ref{Hamiltonian1}) as  

\begin{eqnarray} 
\label{eq:H_ex2}
\hat{H} &=&  \frac{\vec{p}^2}{2\mu} + \frac{\mu \omega_\mathrm{c}^2 \vec{r}^2 }{8} + V(\vec{r}) +    \frac{\nu \omega_\mathrm{c} }{2} [\mathbf{r} \times \mathbf{p}]_z\nonumber\\
&+&\frac{e B}{M} [\vec{r} \times \mathbf{P}]_z  + \frac{\vec{P}^2}{2M}
\end{eqnarray}
with 
\begin{equation}
  \label{defs}
M=m_e+m_h, \quad \mu=\frac{m_\mathrm{e} m_\mathrm{e}}{m_\mathrm{e}+m_\mathrm{h}}, \quad \nu=\frac{m_\mathrm{e}-m_\mathrm{h}}{m_\mathrm{e}+m_\mathrm{h}}. 
\end{equation}

We have also introduced the cyclotron frequency $\omega_c=eB/\mu$.
We restrict ourselves only to the optically active excitons, i.e. excitons with zero center of mass momentum $\vec{P}=0$. Besides, the fourth term is determined by the orbital momentum for the relative motion $\hat{L}_z=[\vec{r}\times\vec {p}]_z$, which is a good quantum number $\hat{L}\to l $. It describes
the orbital Zeeman shift of excitonic levels and is nonzero only for non-s-wave states.




The excitonic states are shaped by the interplay between the attractive Rytova-Keldysh potential and the harmonic confinement due to the magnetic filed.
They smoothly evolve from the 2D hydrogenic-like series to the equidistant set of Landau levels.
It is convenient to calculate the magnetic field dependence of excitonic states using the momentum space representation, i.e., $\psi(\vec{r}) \to \psi_{\bf p}$. The eigenvalue problem with Hamiltonian (\ref{eq:H_ex2}) transforms to the integro-differential equation given by 
\begin{equation}
\label{eq:H_ex_p}
\left[
    \frac{\vec{p}^2}{2 \mu} - \frac{\mu \omega^2_c}{8} \nabla_\vec{p}^2 +  \frac{l \nu  \omega_\mathrm{c} }{2}  \right]\psi_\vec{p} + 
\sum_{\vec{p}'} V_{\vec{p}-\vec{p}'} {\psi_{\vec{p}'}} = \epsilon \psi_\vec{p},
\end{equation}
where  $V_{\bf p}=
-2 \pi e^2/\varepsilon p (1 + r_0 p)$ is the Fourier transform of the Rytova-Keldysh potential. It can be further simplified if we split the orbital part of the wave function $\psi_{\bf p}=
\psi_p^l e^{il\theta_p}$ and rescale its radial part as $\psi_p^l=\chi_p^l/\sqrt{p}$, allowing the radial Hamiltonian to be written in a form that is strictly Hermitian. The resulting eigenvalue problem is given by 
\begin{eqnarray} 
\label{eq:H_ex_her}
&&\left[
  \frac{p^2}{2 \mu} + \frac{\mu \omega_\mathrm{c}^2}{8} \left ( -\frac{d^2}{dp^2}  + \frac{l^2-1/4}{p^2} \right) + \frac{l \nu \omega_\mathrm{c} }{2}  \right]\chi_p^l\nonumber\\
&&+ 
\int_0^\infty\frac{ \sqrt{p\;p'} dp'}{2\pi}\;  V_{l}(p,p') \chi_{p'}^l =\epsilon \chi_p^l \ .
\end{eqnarray}
Here $V_{l}(p,p')=\langle V_{\vec{p}-\vec{p}'} e^{i l \varphi } \rangle_{\varphi}$ is the multipole moment of interactions and involves averaging over the relative polar angle $\varphi$ between momenta $\vec{p}$ and $\vec{p}'$. We solve Eq.(\ref{eq:H_ex_her})  numerically using the linear algebra LAPACK package. Further details of these calculations are presented in Appendix \ref{numerics}.

\section{s-wave Excitons} \label{sec:2}
\begin{figure} [h!t]
	\includegraphics[width=0.36\textwidth]{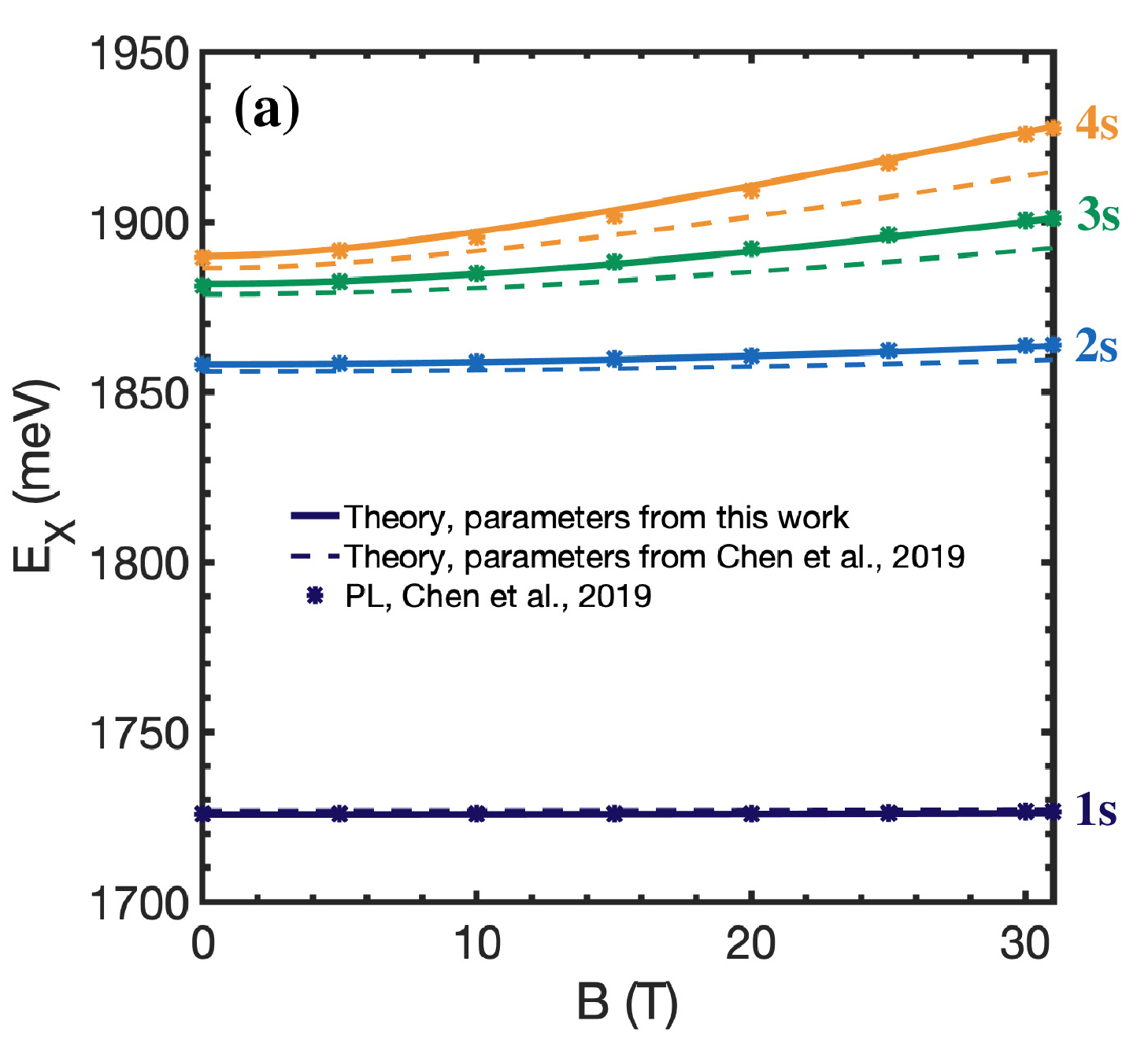}
	\hspace*{0.6cm}\includegraphics[width=0.36\textwidth]{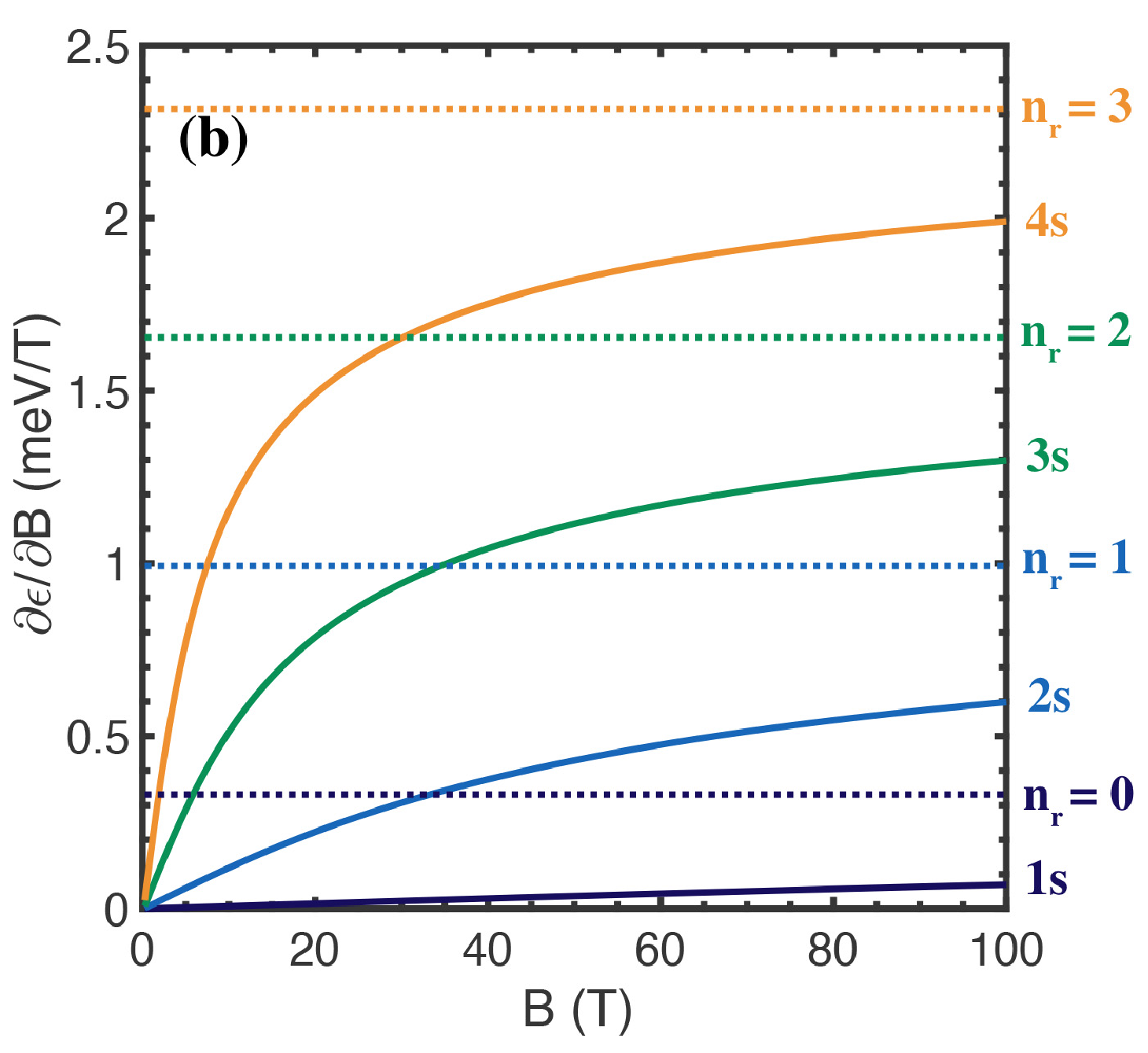}
	\caption{(a) The magnetic field dependence of s-wave excitonic states in WSe$_2$ monolayer encapsulated in hBN. Our theoretical curves (solid lines) are an excellent match to the experimental data from Ref.~\cite{chen_luminescent_2019} (asterisks). Dashed lines show the theory  with the set of parameters obtained by fitting~\cite{chen_luminescent_2019} the ground excitonic state with the perturbative expression, Eq.(\ref{eq:H_simple}).
		(b) The first derivative of excitonic energies $E_{n}$ (solid lines)  and their asymptotic values  $E^\infty_{n}=\omega_\mathrm{c} (n_r+1/2)$ (dashed lines) with respect to the magnetic field. The 1s state remains in the parabolic regime while the Rydberg states indicate transition to the quantized harmonic regime even at experimentally realistic $B$ field.}
	\label{fig:fit}
\end{figure}
We wish to compare the measured magnetic field dependence of energies $E_\mathrm{X}$ for the s-wave excitonic resonances~\cite{chen_luminescent_2019} with the computed excitonic energies $\epsilon$. They are connected as $E_{\mathrm{X}}=\Delta+\epsilon$ where $\Delta\approx 1.9$ eV is the band gap in the TMD monolayer.   
Hence we have 3 fitting parameters: the reduced electron-hole mass $\mu$, the dielectric constant $\varepsilon$, and the
screening length $r_0$. 
As presented in panel a of Fig.~\ref{fig:fit}, our theoretical curves (solid lines) provide an excellent fit of the experimental data (asterisks) for all resolved excitonic states (1s - 4s)\footnote{We reiterate that the experimental points
in Fig.~\ref{fig:fit} correspond to that of 
Fig.~\ref{fig:exp}, yet averaged over $+B$ and $-B$. This averaging removes the linear valley Zeeman effect}.
The fitting parameters are 
\begin{eqnarray}
  \label{fp}
  \mu = 0.175 m_0 \ ,  \ \ \ \varepsilon =3.9 \ , \ \ \ r_0 = 4.5 nm/\varepsilon \ .
\end{eqnarray}  
We make special mention of our fitted value $\varepsilon = 3.9 $, which is within the expected range for hBN~\cite{dean_boron_2010}. With these parameters, the ground state binding energy at zero magnetic field is found to $|\epsilon_{1s}|=173$meV.


The original experimental  work~\cite{chen_luminescent_2019}
has used the simple parabolic approximation (\ref{eq:H_simple}) for the
diamagnetic shift. This has led to a different set of parameters: 
$\mu = 0.22$ m$_0$, $\varepsilon =4.5$, and $r_0 = 4.5$ nm/$\varepsilon$.
The theoretical curves (dashed lines in Fig.~\ref{fig:fit}a) calculated with
this set of parameters fit the data well only for the ground excitonic state,
but the discrepancy for the excited excitonic states is evident.
Our conclusion aligns well with a recent unpublished work \footnote{David de la Fuente Pico, Jesper Levinson, Meera M. Parish and Francesca Maria Marchetti, private communication.}.

\begin{figure} [ht!]
  \includegraphics[width=0.36\textwidth]{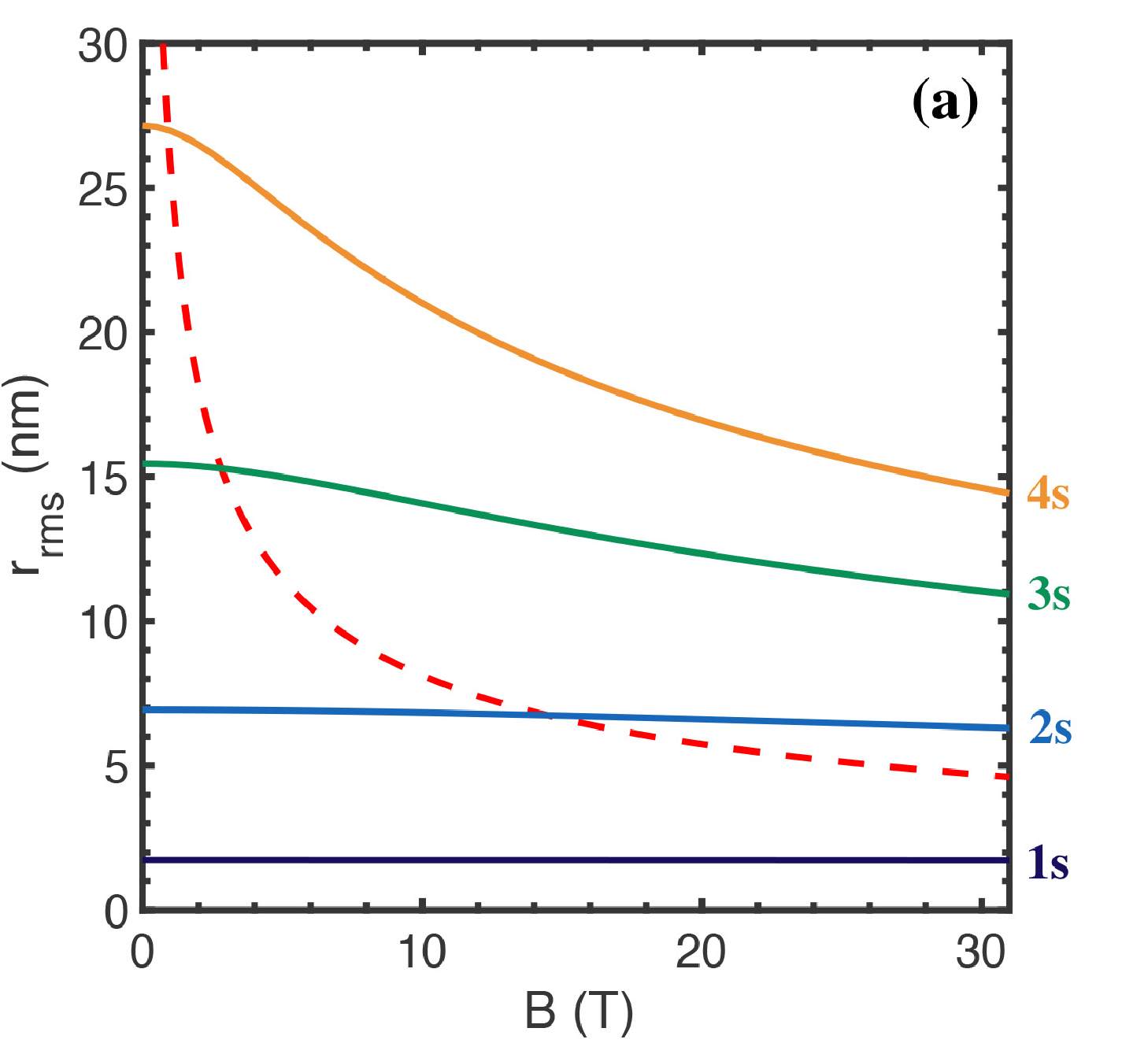}
\includegraphics[width=0.36\textwidth]{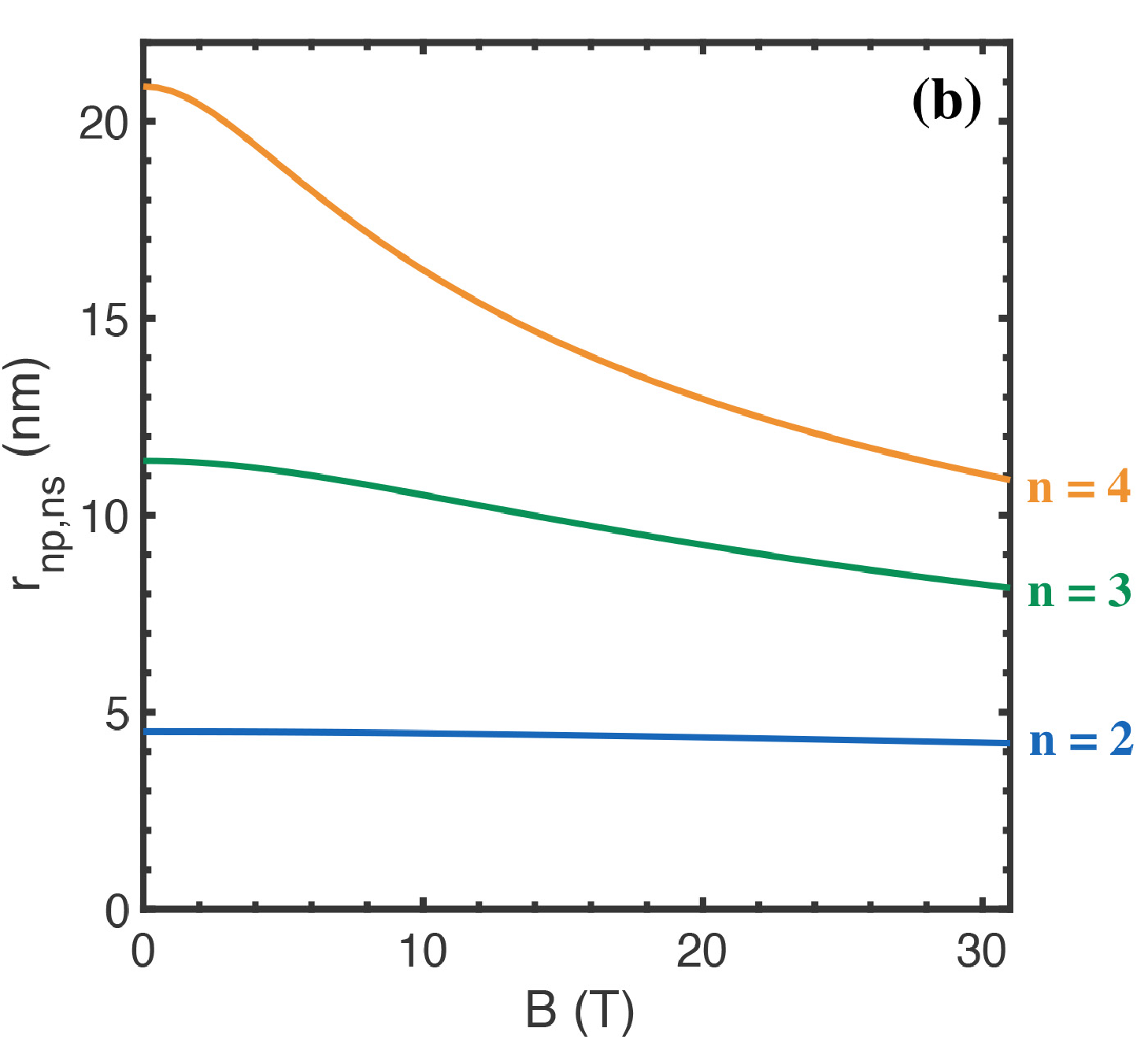}
\caption{(a) Root mean square radius of  1s, 2s, 3s, and 4s
  excitons as a function  of magnetic field. The dashed red line is
  the magnetic length
  $l_B = \sqrt{e/B}$.
  (b) Electric dipole radial matrix element $r_{np,ns}$ for
  $n=2$, $n=3$ and $n=4$ as a function of magnetic field.
}
\label{fig:rr}
\end{figure} 
The behaviour of the eigenenergies of the Rydberg states strays from the parabolic diamagnetic approximation and it can be deduced from Eq.\eqref{eq:H_ex_her} that in the large $B$ limit, $B\to \infty$, the eigenergies approach those of 2D Landau levels.
In this case the eigenenergy is quantized in radial quantum number $n_r$ and angular momentum $l$ as
$\epsilon_{n_r,l} = \omega_c\left(2n_r+|l|+\nu l+1\right)/2$ and is therefore linear in magnetic field $B$. The radial quantum number $n_r$ is related to the principle quantum number $n$ as $n_r=n-|l|-1$.
In panel b of Fig.~\ref{fig:fit} we present the calculated  first derivative of the excitonic energy calculated from Eq.\eqref{eq:H_ex_her} with respect to magnetic field. Solid lines in Fig.~\ref{fig:fit}b correspond to solid lines in Fig.~\ref{fig:fit}a. Dashed lines in Fig.~\ref{fig:fit}b correspond to asymptotic values of the derivatives ($B\to \infty$) determined by Landau levels in 2D. While in the experimental region $0 < B <30$T the derivative of the 1s state is practically linear in $B$ in accordance with perturbation theory, for higher states deviations from linearity are evident. Thus it is clear that the Rydberg states transition from the diamagnetic regime towards the Landau level regime at experimentally realistic $B$, rendering the quadratic diamagnetic approximation Eq.\eqref{eq:H_simple} invalid for the Rydberg excitons.

The region of validity of the diamagnetic approximation is also illustrated by considering the size of the exciton, or the root mean square (rms) radius $r_{rms}$. The size of the exciton may be calculated directly from the wavefunction.
\begin{equation}
    \langle|r^2|\rangle = \langle\psi|r^2|\psi\rangle = \int_0^\infty \left( \left(\frac{\partial\psi_p^l}{\partial p}\right)^2 p +(\psi_p^l)^2\frac{l^2}{p} \right)\frac{dp}{2\pi}
\end{equation}
The exciton size $r_{rms}=\sqrt{\langle\psi| r^2|\psi\rangle}$ is plotted  against magnetic field in panel a of Fig.~\ref{fig:rr}. The 1s exciton radius is $\sim2$ nm and remains smaller than the magnetic length $l_B = \sqrt{ e/B}$ over the entire range of magnetic field, however this is not true for the Rydberg excitons.
Thus, to accurately model the Rydberg states over this range of $B$ the exact Hamiltonian \eqref{eq:H_ex_her} is required.

\section{Detection of p-wave excitons via mixing due to in-plane electric field} 
\label{sec:EE}
\begin{figure}[ht!] 
	\centering
	\includegraphics[width=0.36\textwidth]{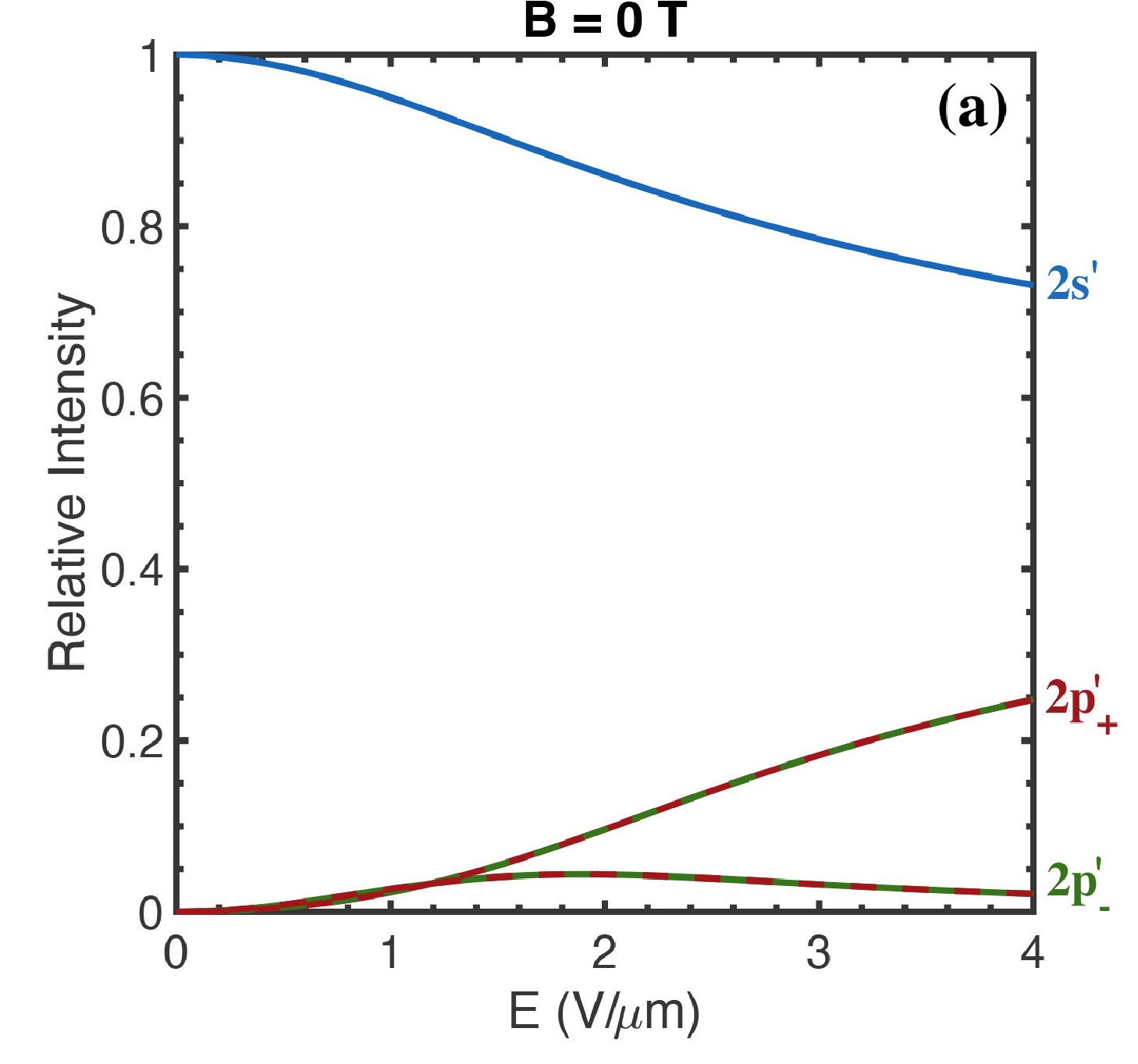}  
	\includegraphics[width=0.36\textwidth]{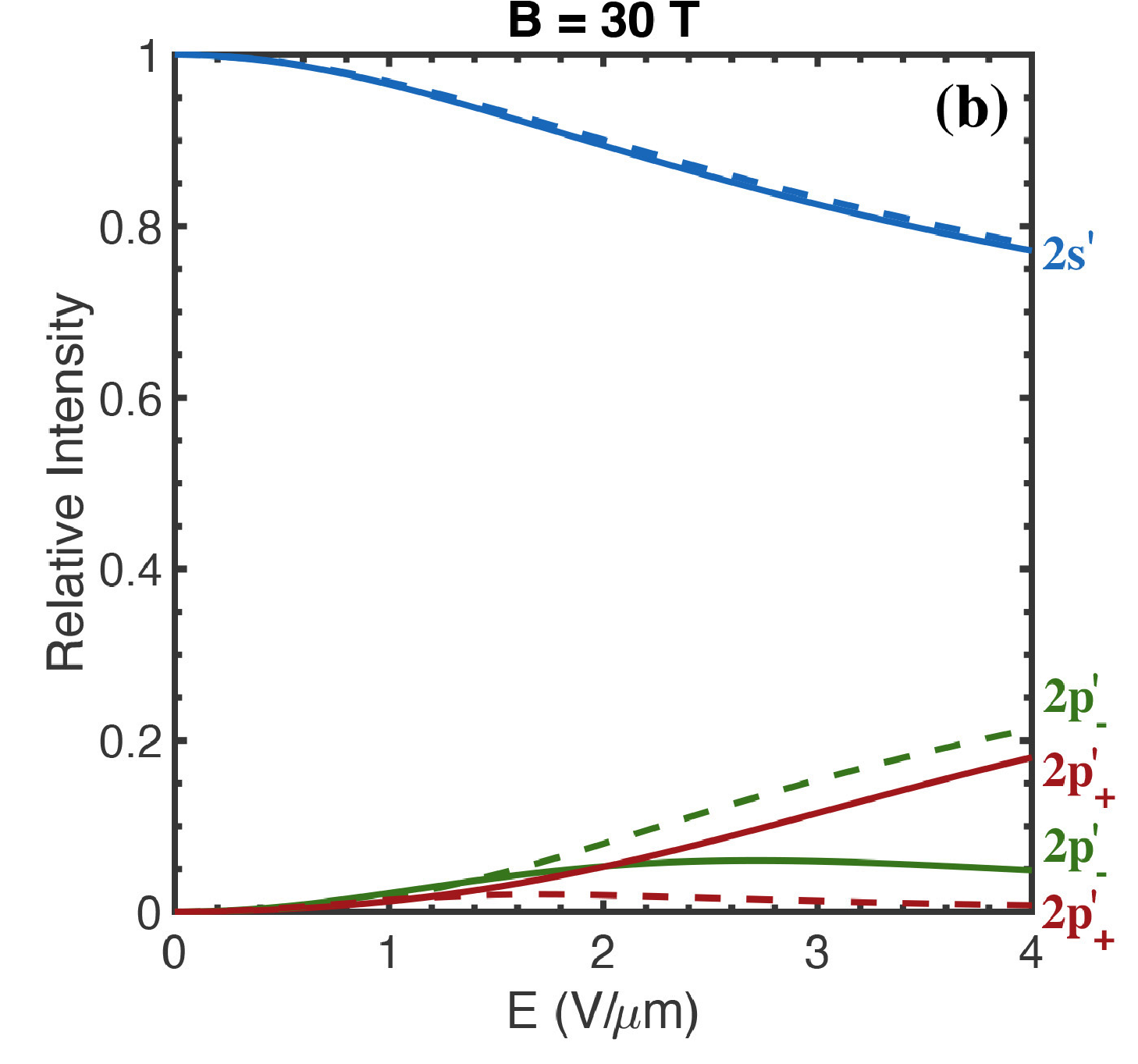}  
	\caption{Relative excitation probabilities of $2s^\prime$, $2p_+^\prime$,
		and $2p_-^\prime$ exciton states in different valleys of WSe$_2$ versus
		external in-plane
		electric field E. Linear polarization of the exciting laser is assumed.
		Panel a corresponds to zero out-of-plane magnetic field, $B=0$.
		Panel b corresponds to  $B=30$T.
		Solid lines correspond to $\tau=+1$ valley and dashed lines
		correspond to $\tau=-1$ valley.
		For B=0 the $p_\pm^\prime$ curves in different valleys are identical,
		but what is the $p_+^\prime$ curve  for one valley is the $p_-^\prime$ curve for
		the other valley. For $B\ne 0$ all the curves are different.
	}
	\label{IntensityE}
\end{figure}
The physics of p-wave exciton states is richer than that of s-wave states.
It includes the valley-orbital effect (the splitting of $l=\pm 1$ states
without magnetic field) and the orbital Zeeman effect (the splitting of
$l=\pm 1$ states in magnetic field).
Unfortunately the spectral weight of direct photoexcitation of a p-wave
exciton calculated in
 Appendix~\ref{app:mixing} is too small for direct
 detection, $\sim10^{-3}$ in relative units compared with the s-wave state.
 However, one can mix s and p states
 by applying in-plane electric field and hence excite the p state due to admixture of the s state.

\begin{figure}[ht!]
\includegraphics[width=0.36\textwidth]{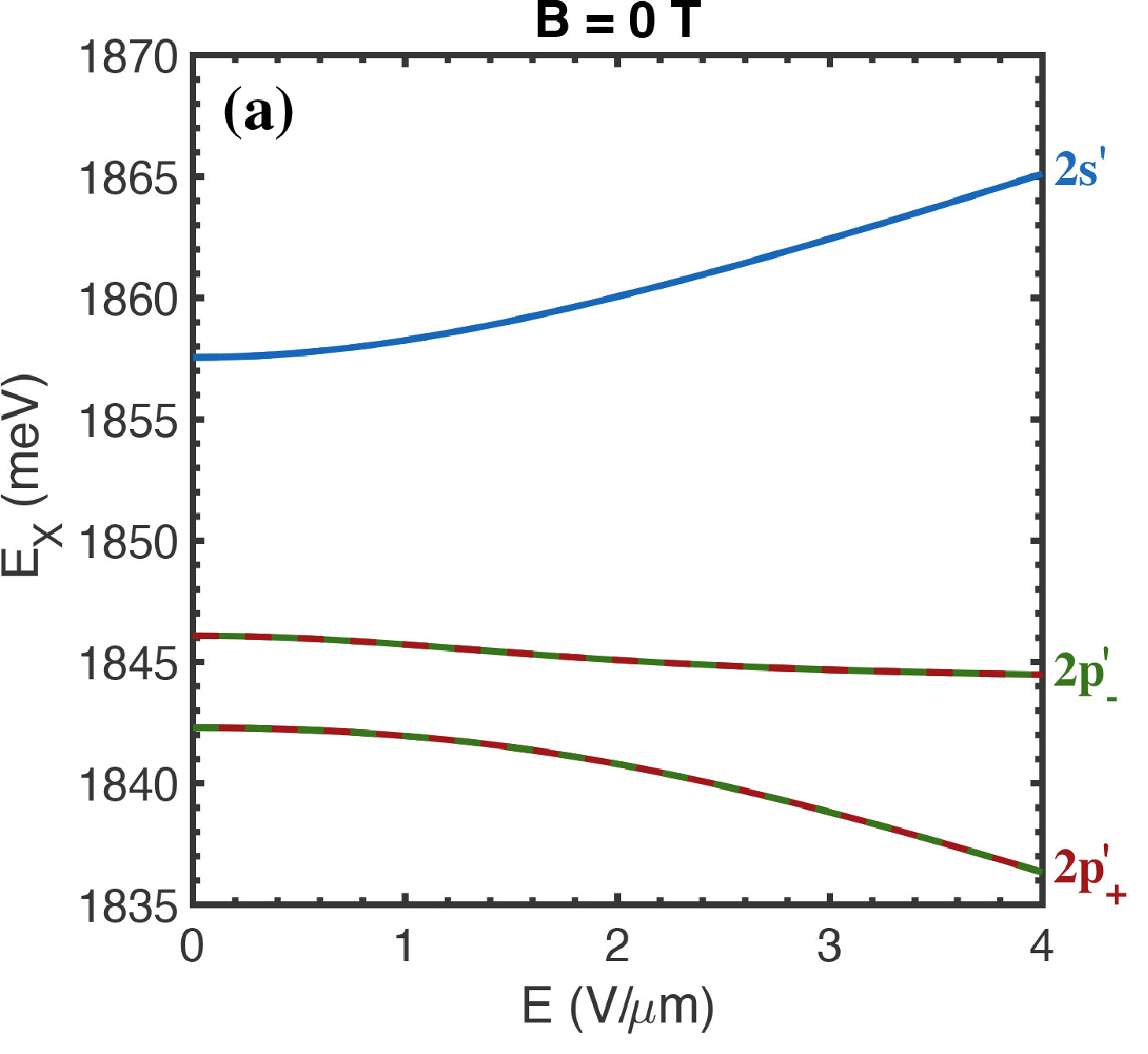}
\includegraphics[width=0.36\textwidth]{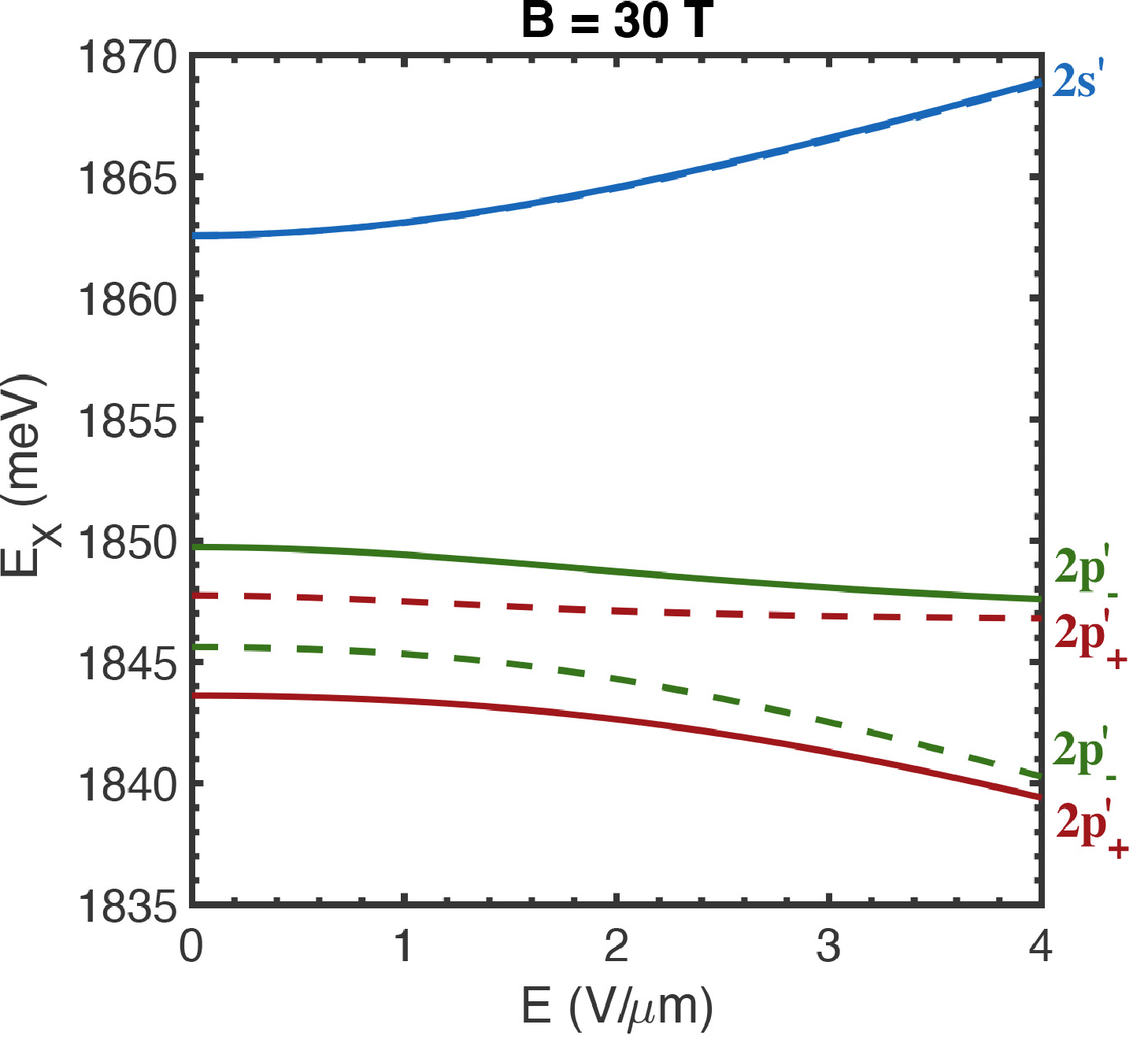}
\caption{Eigenenergies of $2s^\prime$, $2p_+^\prime$, and $2p_-^\prime$ exciton
  states in different valleys of WSe$_2$ versus external in-plane electric field E.
 Panel a corresponds to zero out-of-plane magnetic field, $B=0$.
  Panel b corresponds to  $B=30$T.
  Solid lines correspond to $\tau=+1$ valley and dashed lines
  correspond to $\tau=-1$ valley.
}
\label{eigenE}
\end{figure}
There is a recent experiment where the p-wave states have been observed
using this method~\cite{zhu_-plane_2023}.
However, in experiment~\cite{zhu_-plane_2023} the electric field was
relatively strong
and an interpretation in terms of simple electron-hole bound states (excitons)
is questionable. The major part of Ref.~\cite{zhu_-plane_2023} data
corresponds to events above the exciton ionisation limit where the
observed resonances correspond to doorway scattering
states~\footnote{O. P. Sushkov, J. N. Engdahl and D. K. Efimkin,
to be published}.
  Contrary to this in the present work we consider a relatively weak
  electric field when the description in terms of bound excition states
  is correct.

  Energies of np-states with or without magnetic field
  can be calculated using Eq.(\ref{eq:H_ex_her}). The np level is always
  slightly lower than the ns-level. However, Eq.(\ref{eq:H_ex_her}) is missing
  an important effect: the valley-orbital level splitting between the states
  $p_+$ and $p_-$ that correspond to angular momentum $l=+1$ and $l=-1$, where this splitting exists even at B=0.
  Using the language of Dirac equation one can say that Eq.(\ref{eq:H_ex_her})
  corresponds to the ``non-relativistic approximation'' to the Dirac equation
that at $B=0$ gives degenerate $p_\pm$ states,
and the $p_+$ - $p_-$ splitting is the 1st ``relativistic'' correction
to this equation,
  The splitting is discussed in Appendix~\ref{app:pseudo}, it is small,
  from only a few to several meV, and if it is positive in one valley,
  $\Delta\epsilon_\tau=\epsilon_--\epsilon_+ >0$, it is negative in the other
  valley.

  Having in mind that energies of $ns$, $np_+$ and $np_-$ states with the same n
  are close and  assuming that the electric field $E=E_x$ is not too strong we can restrict
  our analysis to a 3-level approximation with the effective Hamiltonian
  \begin{align}
    \label{mixing_H}
    \hat{H} 
    = \begin{bmatrix}
            \epsilon_{s,0}&
            v_{s,p_-} & v_{s,p_+}\\
            v_{p_-,s}&
            \epsilon_{p_-,0}+\Delta\epsilon_\tau/2 &0\\
             v_{p_+,s}&0&\epsilon_{p_+,0}-\Delta\epsilon_\tau/2
     \end{bmatrix}
\end{align}
 Here $\epsilon_{a,0}$ is the eigenenergy of the state with
 angular momentum $l_a$ calculated with Eq.(\ref{eq:H_ex_her}).
 $\Delta\epsilon_\tau$ is the valley-orbital splitting, and
 $v_{a,b}=eE_x x_{a,b}$,  where 
 \begin{equation}
   \label{eq:radial_element}
   x_{a,b} = \langle a|x|b\rangle =\int \psi_{a}^* e^{-il_a\varphi} \frac{\partial}{\partial p_x}\left(\psi_{b} e^{il_b\varphi}\right) \frac{pdp}{2\pi}
=r_{ps}/2 \ .
 \end{equation}
 In panel b of Fig.~\ref{fig:rr} we present plots of   the electric dipole
 radial matrix element $r_{np,ns}$ for $n=2,3,4$ as a function of magnetic field.
 
 Before diagonalization of \eqref{mixing_H} the energies
  $\epsilon_{s,0}$, $\epsilon_{p_+,0}$ and $\epsilon_{p_-,0}$
must be calculated from Eq.(\ref{eq:H_ex_her}). From the fit of the
 s-wave states we know all the parameters of this equation except for $\nu$,
 the particle-hole asymmetry term defined in (\ref{defs}).
 Here we take $\nu=-0.1$, which is obtained from the effective electron and
 hole masses determined from DFT calculations in Ref.~\cite{kormanyos_k_2015}.  However, armed with our formalism, we suggest that analysis of future experiments can establish $\nu$, providing vital information on the underlying band structure. 
 The valley-orbital splitting $\Delta\epsilon_\tau$ is discussed and
 calculated in Appendix~\ref{app:pseudo}. In principle it depends on
 magnetic field, but for n=2 this dependence is negligible
 over the range of magnetic field that we consider.
 The splitting depends on the velocity in the effective Dirac equation,
see Appendix~\ref{app:pseudo}.
 We approximate the velocity from the band gap and the effective mass as $v^2\approx\Delta/4\mu$, which results in a valley-orbital splitting
 $\Delta\epsilon_\tau \approx\pm 3.8$meV.
 Finally we reiterate the point that we already have made in Section~\ref{background}:
 the linear in B valley Zeeman effect that is not exciton specific and can be
 determined from the 1s state data is subtracted from all our results.
 After this preparatory procedure, diagonalization of Eq.\eqref{mixing_H}
 is straightforward. We use nomenclature $s^\prime$, $p_+^\prime$ and $p_-^\prime$ to denote
 the states that originate from $s$, $p_+$ and $p_-$ at $E=B=0$.
 The photoexcitation probability is proportional to the weight of the bare
 s-wave state in the corresponding wave function.
 In Fig.\ref{IntensityE} we plot the probabilities versus electric field
 for n=2 states.
 The probabilities are normalized  such that $P_{2s}=1$ at $E=0$.
As previosuly mentioned, the approach used in the present work fails at strong electric fields. This is due to the effective potential of the Keldysh potential combined with the strong electric field failing to permit exciton bound states. A separate analysis of strong electric fields~\footnote{O. P. Sushkov, J. N. Engdahl and D. K. Efimkin, to be published} shows that for n=2 states in WSe$_2$ the
 approach employed in the present work is valid up to $E=3-4$V/$\mu$m,
above this field  bound n=2 exciton states do not exist.
  This is where we terminate our plots in Fig.\ref{IntensityE}.
 Here the probability of the $p^\prime$-state excitation can be up to 20\%
 of that of the s-wave.
 The relative probabilities in Fig.\ref{IntensityE} assume linear polarization
 of the excitation laser that does not separately select single valleys.
 Panel a in  Fig.\ref{IntensityE} displayes probabilities at B=0.
 Plots for different  valleys are identical, but what is the $p_+^\prime$ curve
 for one valley is the $p_-^\prime$ curve for the other valley.
 Panel b in  Fig.\ref{IntensityE} displayes probabilities at B=30T.

 In Fig.\ref{eigenE} we plot energies of $2s^\prime$, $2p_+^\prime$, and $2p_-^\prime$ states versus external in-plane electric field E for two values of
 magnetic field, $B=0$ and $B=30$T.
 These plots depend on the valley-orbital energy splitting between the
 $p_\pm$ levels, $\Delta\epsilon_\tau$,  and also on the mass asymmetry
 parameter $\nu$-term in Eq.(\ref{eq:H_ex2}).
 In Fig.\ref{eigenE}  we assume values that follow from DFT
 calculations\cite{kormanyos_k_2015} and indirect analysis of ARPES data\cite{nguyen_visualizing_2019}.
 Measurements of these energy levels would shed light on true values of these
 parameters.

\section{Conclusion} \label{sec:3}
  We have shown that the perturbation theory approximation for diamagnetic shift
  of exciton energy levels is not sufficient to describe existing data.
  We develop the theory for strong magnetic field, reanalyse the data for 
  monolayer  WSe$_2$, and arrive to the set of exciton parameters presented in
  Eq.(\ref{fp}).  This set is different from what was known in literature, although the reduced mass is similar to that obtained from DFT calculations.
  
  Only s-wave excitons are visible in photoluminescence, the spectral weight
  of p-wave states is too small at just $\sim 10^{-3}$ relative to the s-wave.
  We propose to study p-wave states by mixing them with s-wave states by application of 
  external in-plane electric field and show that a moderate  electric
  field $E\approx 3V/\mu m$ results in the intensity about 20\%  relative
  to s-wave.
  We calculate energy levels in combined electric and magnetic fields and
  demonstrate that our proposal opens a path to experimentally study the
  valley-orbital
  level splitting and the electron-hole mass asymmetry.

\section*{ Acknowledgements} 
We acknowledge discussions with Zeb Krix, Francesca Marchetti and Jesper Levinson. This work was supported by the Australian
Research Council Centre of Excellence in Future Low-
Energy Electronics Technologies (CE170100039). 

\appendix

\section{Reduction of the Dirac like  Hamiltonian to the low energy Hamiltonian,
valley Zeeman g-factor.}
\label{app:zeeman}
Start from the 2D Dirac Hamiltonian, Eq.(\ref{HC1}), for a particle in
1L TMD~\cite{rostami_effective_2013}, truncated to order quadratic in momentum,
${\bf q}=-i\partial_r$. Parameters $\alpha$ and $\beta$ define particle-hole mass asymmetry, $\gamma \approx 1$ determines spin splitting of the conduction band ($\gamma = 1$ gives only splitting of the valence band) and $\lambda$ determines the magnitude of this spin splitting. $\Delta_0$ is the band gap in the absence of spin-orbit coupling and $v$ is Fermi-Dirac velocity. Note that in this notation the valley and spin indices are $\tau = \pm 1$ and $s = \pm 1$. 
\begin{equation}
\label{HC1}
    \hat{H} = \frac{\Delta_0}{2} \sigma_z + \lambda \tau s \frac{\gamma - \sigma_z}{2} + v \mathbf{q} \cdot \mathbf{\sigma_\tau} + \frac{|\mathbf{q}|^2}{4 m_0}(\alpha + \beta \sigma_z) 
\end{equation}
We can write this as a 2x2 eigensystem, where $q_\pm = \tau q_x \pm iq_y$, $\sigma_\tau = (\tau\sigma_x,\sigma_y)$, $\lambda_1 = (\lambda \tau s/2)(\gamma - 1)$ and $\lambda_2 = (\lambda \tau s/2)(\gamma + 1)$. 

\begin{multline}
\label{H_zeeman}
    \hat{H} 
    = \\\begin{bmatrix}
            \frac{\Delta_0}{2} + \lambda_1 + \frac{q^2}{4m_0}(\alpha+\beta) &
            vq_- \\
            vq_+ &
            \frac{-\Delta_0}{2} + \lambda_2 + \frac{q^2}{4m_0}(\alpha-\beta)
         \end{bmatrix}
\end{multline}
\begin{align}
    \hat{H} \begin{bmatrix}
           \psi_1 \\
           \psi_2
         \end{bmatrix} = \epsilon \begin{bmatrix}
           \psi_1 \\
           \psi_2
         \end{bmatrix}
\end{align}

Excitons that we consider have binding energy much smaller than the band gap.
For Dirac equation this corresponds to non-relativistic limit.
Hence we follow the standard procedure of derivation of Pauli
equation from Dirac equation~\cite{beresteckij_quantum_2008}.
We  explicitly write out the upper and lower component  equations, substituting $\psi_2$ and $\psi_1$ into the equations respectively and then taking the
limit $\epsilon \approx \pm \Delta$. For the positive energy energy solution
$\psi_2 \ll \psi_1$ and for the negative energy energy solution
$\psi_1 \ll \psi_2$. Hence we arrive to the following energies.
\begin{multline}
    \epsilon_1 = \frac{\Delta_0}{2} + \frac{\lambda \tau s_1}{2} (\gamma -1) + \frac{q_1^2}{4m_0}(\alpha + \beta) + \frac{v^2 q_1^2}{\Delta_0 - \frac{\lambda \tau s_1}{2}(\gamma + 1)}
    \\
    \epsilon_2 = \frac{-\Delta_0}{2} + \frac{\lambda \tau s_2}{2} (\gamma +1) + \frac{q_2^2}{4m_0}(\alpha - \beta) - \frac{v^2 q_2^2}{\Delta_0 + \frac{\lambda \tau s_2}{2}(\gamma - 1)}    
\end{multline}
Now, to go to the hole description instead of the electron description,
we perform charge conjugation on the negative energy solution, where $\mathbf{q_2} \to \mathbf{-q_2}$, $s_2 \to -s_2$ and $\epsilon_2 \to -\epsilon_{2,+}$.
\begin{eqnarray}
  \label{ae2}
  \epsilon_{2} &=& \frac{\Delta_0}{2} + \frac{\lambda \tau s_2}{2} (\gamma +1) - \frac{q_2^2}{4m_0}(\alpha - \beta)\nonumber\\
&+& \frac{v^2 q_2^2}{\Delta_0- \frac{\lambda \tau s_2}{2}(\gamma - 1)}    
\end{eqnarray}

In presence of a magnetic field we first perform gauge replacement
$\mathbf{q} \to \mathbf{q} + e\mathbf{A}$ in the Hamiltonian (\ref{HC1})
and then repeat the above procedure. Hence, we arrive at the following. 
\begin{multline}
\label{ae3}
       \epsilon_1 = \frac{\Delta_0}{2} + \frac{\lambda \tau s_1}{2} (\gamma -1) + \frac{|\mathbf{q_1} + e\mathbf{A_1}|^2}{4m_0}(\alpha + \beta)\\ + \frac{v^2((\mathbf{q_1} + e\mathbf{A_1})^2 + \tau eB)}{\Delta_0 - \frac{\lambda \tau s_1}{2}(\gamma + 1)}
    \\
    \epsilon_{2} = \frac{\Delta_0}{2} + \frac{\lambda \tau s_2}{2} (\gamma +1) - \frac{|\mathbf{q_2} - e\mathbf{A_2}|^2}{4m_0}(\alpha - \beta)\\ + \frac{v^2((\mathbf{q_2} - e\mathbf{A_2})^2 - \tau eB)}{\Delta_0 - \frac{\lambda \tau s_2}{2}(\gamma - 1)}   
\end{multline}
These equation represent effective single particle Hamiltonians for electron and hole.

Finally, let us consider an excitation of electron from the valence
band to the conduction band via E1 optical transition. The excitation
creates an electron and a hole.
The spin of the electron is $s_e = s_1 = s$ and the spin of the hole is $s_h = -s_1 = -s$. The Hamiltonian of the system is the combined Hamiltonian of the electron and hole. The linear Zeeman term may be denoted as $\hat{H}_Z$. 
\begin{eqnarray}
    \hat{H}_Z &=& -v^2\tau eB\left(\frac{1}{\Delta_0 + \frac{\lambda \tau s}{2}(\gamma - 1)} - \frac{1}{\Delta_0 - \frac{\lambda \tau s}{2}(\gamma + 1)} \right)
    \nonumber\\
    &\approx&-\frac{v^2\tau eB\lambda}{\Delta_0(\Delta_0-\lambda)}
\end{eqnarray}
Here in the second line we take the small conduction band spin-orbit-splitting limit $\gamma \approx 1$.
The $v^2$-terms in Eqs.(\ref{ae3}) contribute to the effective masses of electron and hole. It is easy to check that the following relation is
valid.
\begin{equation}
  \label{v2}
  \frac{v^2}{\Delta_0(\Delta_0-\lambda)} = \frac{1}
       {2m_0(2\Delta_0-\lambda)}\left(\frac{m_0}{\mu} +\beta \right).
\end{equation}
Note also that in the language of Eq.~\eqref{H_zeeman} $\Delta_0$ refers to the average band gap for the two spin split band pairs, meaning for A-excitons the band gap is $\Delta=\Delta_0-\lambda$. Thus we may rewrite the valley Zeeman
term as
\begin{equation}
\label{eq_zeeman}
      {\hat H}_Z \simeq -\tau\frac{\mu_BB\lambda}{2\Delta+\lambda}
      \left(\frac{m_0}{\mu}+\beta\right)= -g\tau \mu_BB
\end{equation}
The value of the spin-orbit constant according to
DFT calculations\cite{kormanyos_k_2015} and experimental
measurements~\cite{nguyen_visualizing_2019} is  about $\lambda\approx 0.24$eV.
The value of the reduced mass $\mu$ is known, Eq.(\ref{fp}).
The value of the Fermi-Dirac velocity  extracted
from ARPES data for a similar hBN/WSe$_2$/hBN device \cite{nguyen_visualizing_2019} is $v\approx 0.4$eVnm.
Comparing this with Eq.(\ref{v2}) one finds
that the $\beta$-term in Eqs.(\ref{v2}),(\ref{eq_zeeman}) is small compared to
$m_0/\mu$. 
Hence, using Eq.(\ref{eq_zeeman}) we find the theoretical prediction for
the valley Zeeman g-factor, $g\approx 0.34$. On the other hand experimental
data from
Ref.~\cite{chen_luminescent_2019} presented in our Fig.\ref{fig:exp} gives
$g=2.1$, a dramatic disagreement between the theory and the experiment.
Such a disagreement for the valley-Zeeman g-factors in TMD materials is known
in literature~\cite{wozniak_exciton_2020,rybkovskiy_atomically_2017}.
There are also claims in literature~\cite{wozniak_exciton_2020} that account of
multiple bands (up to 200 bands) can bring theory to agreement with experiment.
The dramatic disagreemt is a very interesting problem, but it is beyond the
scope of the current work since it is irrelevant to questions considered here.

\section{Numerical Methods}
\label{numerics}
We solve Eq.~\ref{eq:H_ex_her} through linear algebra methods. 
The angular integration
in the potential in Eq.
~\ref{eq:H_ex_her} is 
performed with  $200$ 
points, so $\Delta\theta=2\pi/200$.
The radial
momentum is discretized with 500 grid
points, the grid step is
$\Delta p = 0.01/a_B$, where $a_B\approx 1.2$ nm is effective Bohr radius.
Hence the Hamiltonian 
in  Eq. ~\ref{eq:H_ex_her} 
is a $500\times 500$ matrix.
There is a small pitfall with
this method, Eq.(\ref{eq:H_ex_her}) is singular for s-wave states
at $p\to 0$. If Eq.\eqref{eq:H_ex_her} for $l=0$ is written symbolically as 
\begin{equation} 
\label{eq:H_symbolic}
    \hat{H} = \frac{\vec{p}^2}{2 \mu} + \frac{e^2 B^2 r^2}{8 \mu c^2 }  + V(\vec{r})
\end{equation}
is is clear that the singularity is due to
$r^2$ divergence  of (\ref{eq:H_symbolic}) at large r.
Hence we have to be careful with discretization at small p. We resolve this by adding a regularization factor to the matrix elements at the smallest momentum. A simple test
that the discretization is correct is that eigenenergies and eigenfunctions 
of Eq.(\ref{eq:H_ex_her}) at $V=0$ coincide with that of Eq.(\ref{eq:H_symbolic}) obtained by the  conventional analytic method.

\section{s-wave and p-wave relative photoexcitatin spectral weights}
\label{app:mixing}
Firstly, consider the Hamiltonian for an insulator without electromagnetic field.
In the Hamiltonian (\ref{HC1})
we can disregard the second and the fourth terms, but we 
must add the 
trigonal warping ~\cite{rostami_effective_2013} $t$-term that is
responsible for excitation
of a p-wave state.
Hence we arrive at
\begin{align}
    \hat{H} = &\frac{\Delta}{2} \sigma_z + v \mathbf{q} \cdot \mathbf{\sigma_\tau} + t\mathbf{q}\cdot \sigma_\tau^* \sigma_x \mathbf{q}\cdot \sigma_\tau^*
\end{align}
In matrix form it is
\begin{align}
\label{HAA}
    \hat{H} 
    = \begin{bmatrix}
            \frac{\Delta}{2} &
            vq_- + tq_+^2\\
            vq_+ +tq_-^2&
            \frac{-\Delta}{2}
         \end{bmatrix}
\end{align}

It is convenient to work in the gauge where the scalar potential is zero.
Hence the vecor potential potential and the electric field of the
photon are related as ${\bf A}={\bf E}_0/\omega e^{-i\omega t}$
To  be specific we consider light
linearly polarized along x.
\begin{equation}
    A_+ = A_- = \tau A_x = \tau A = \frac{E_0}{\omega} \;.\; E = E_0 e^{-i\omega t}
\end{equation}
Interaction with photon
arises from the standard gauge replacement
${\bf q}\to {\bf q}+e{\bf A}$ in the 
Hamiltonian (\ref{HAA}).
Here ${\bf A}$ is vector
potential of the photon.
This give the following
interaction Hamiltonian
linear in ${\bf A}$.
\begin{eqnarray}
    \hat{H}_{int} 
    &=& \begin{bmatrix}
            0 &
            evA_- + 2etq_+A_+\\
            evA_+ + 2etq_-A_-&
            0
    \end{bmatrix}\nonumber\\
    &\to& e\tau A \begin{bmatrix}
            0 &
            v + 2tq_+\\
            v + 2tq_-&
            0
         \end{bmatrix}
\end{eqnarray}

The matrix element, M, for the
electron exitation from the valence 
(lower) band to the conduction (upper)
band is
\begin{align}
    \frac{M}{e\tau A} = \begin{bmatrix}
           1 \\
           0
         \end{bmatrix}^\dag \frac{\hat{H}_{int}}{e\tau A}\begin{bmatrix}
           0 \\
           1
         \end{bmatrix}
         = v + 2tq_+
\end{align}
This is written in terms of plane waves. We need to combine the plane wave decomposition to the exciton wave function $\psi_q$
\begin{equation}
    M \to \int M \psi_q \frac{d^2q}{(2\pi)^2}
\end{equation}
At this point, recall $q_+ = \tau q_x + iq_y$ and $\psi_q(q,\phi) = \psi_q^l e^{il\phi}$. 
Below the
superscript $(+)$ and $(-)$ for p-wave refers to $l=+1$ and $l=-1$ respectively.
\begin{equation}
    \frac{M_{s}}{e\tau A} = \int v \psi_{q,s} q \frac{dq}{2\pi}
\end{equation}
\begin{equation}
    \frac{M_{p}^{(+)}}{e\tau A} = \int t \psi_{q,p} q^2 (\tau-1) \frac{dq}{2\pi}
\end{equation}
\begin{equation}
    \frac{M_{p}^{(-)}}{e\tau A} = \int t \psi_{q,p} q^2 (\tau+1) \frac{dq}{2\pi}
\end{equation}
We see that
with linearly polarized light the chirality of the excited p-wave state is valley dependent. The Fermi-Dirac velocity is approximately
$v\approx\sqrt{\Delta/4\mu}\approx 0.45$ eVnm,
see discussion in Appendix \ref{app:zeeman}.
Hence,
taking $t = -0.93\times10^{-2}$ eVnm$^2$ from DFT calculations from Ref.~\cite{kormanyos_k_2015}, we calculate the magnitude of the matrix elements. Here we take $n=2$ such that we consider $2$s and $2$p.
\begin{eqnarray}
&&\left|\frac{M_{2s}}{-e\tau A}\right|  \propto 6.2\times 10^{-2} \nonumber\\
&&\left|\frac{M_{2p}}{-e\tau A}\right| 
    \propto 3.4\times 10^{-3}
\end{eqnarray}
The spectral weight is proportional to $|M^2|$.
Hence the ratio of 2s and 2p spectral weights
is
\begin{equation}
    \frac{w_{0\to 2p}}{w_{0\to 2s}} = \frac{|M_{2p}|^2}{|M_{2s}|^2} \approx 3\times10^{-3}
\end{equation}

Signatures of p-wave excitons are thus faint compared with s-wave excitons in such systems. We find this value is approximately independent of magnetic field strength for the fields considered in this work. 

\section{Valley-orbital splitting}\label{app:pseudo}
To calculate the valley-orbital we use the same method that is used
for calculation of spin-orbit interaction in positronium~\cite{beresteckij_quantum_2008}.
So, first we consider scattering of electron from hole.
The scattering amplitude is given by the  following diagram.
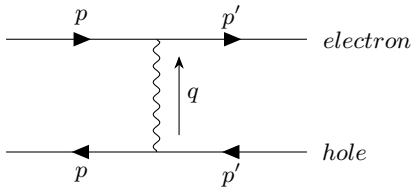
\begin{figure}[ht!]
    \centering
\begin{tikzpicture} 
    \begin{feynman}
            \vertex at (1.0, 1.8)  {\(p\)};
        \vertex at (3.0, 1.8)  {\(p'\)};
             \vertex at (1.0, -0.3)  {\(p\)};
        \vertex at (3.0, -0.3)  {\(p'\)};
                \vertex at (4.8, 1.5)  {\(electron\)};
                \vertex at (4.5, 0.0)  {\(hole\)};
        \vertex at (0.0, 1.5) (a);
        \vertex at (2.0, 1.5) (b);
        \vertex at (4.0, 1.5) (c);
        \vertex at (0.0, 0.0) (d);
        \vertex at (2.0, 0.0) (e);
        \vertex at (4.0, 0.0) (f);
        
    \diagram*{
        (a) -- [fermion] (b) -- [fermion] (c), (f) -- [fermion] (e) -- [fermion] (d), (e) -- [boson,momentum'=\(q\)] (b)
    };
    \end{feynman}
    \end{tikzpicture}
    \caption{Electron-hole scattering  amplitude}
    \label{fig:exchange}
\end{figure}
The amplitude is
\begin{equation}
  \label{M1}
  M = -U_q \langle\Psi^*_{p'}|\Psi_p\rangle_{electron}
  \langle\Psi^*_{p}|\Psi_{p'}\rangle_{hole}.
\end{equation}
Here $U_q=\frac{2\pi e^2}{\varepsilon q(1+r_0q)}$ is the Fourier
transform of  repulsive Keldysh potential and  $\Psi_p$ is
the two component Dirac-like eigenfunction  of electron/hole
discussed in Appendix \ref{app:zeeman}.
\begin{eqnarray}
electron: \ &&    \Psi_{p} \approx \begin{bmatrix}
           1 \\
           \frac{vp_+}{\Delta}
         \end{bmatrix}
    \nonumber\\
 hole: \  \ \ \ \ \ \ &&         \Psi_{p} = \begin{bmatrix}
           -\frac{vp_-}{\Delta} \\
           1
         \end{bmatrix}
\end{eqnarray}
The common sign (-) in Eq.(\ref{M1}) indicates that the interaction is
attractive.
Direct evaluation of (\ref{M1}) gives
\begin{align}
\label{M2}  
    M &= -U_q(1+2\frac{v^2}{\Delta^2}p_-^\prime p_++...)\nonumber \\
    &= -U_q(1+2\frac{v^2}{\Delta^2}(p^2+
    \mathbf{q}\cdot\mathbf{p}-i\tau[\mathbf{p}\times\mathbf{q}]_z)...)
\end{align}
Here ${\bf q}={\bf p}^\prime-{\bf p}$, we keep only the $p_-^\prime p_+
\to -i\tau[\mathbf{p}\times\mathbf{q}]_z$ term
relevant for the valley-orbit interaction.
It is easy to check that this term in (\ref{M2})
is the matrix element
$\langle {\bf p}^\prime|H_{vo}|{\bf p}\rangle$ of the effective valley-orbit
Hamiltonian
\begin{eqnarray}
  \label{Hvo}
  H_{vo}=-2\tau\frac{v^2}{\Delta^2}\nabla U_r\times\mathbf{p},
  \end{eqnarray}
where $U_r$ is repulsive Keldysh potential in coordinate representation.
Noting that $\nabla U_r\times\mathbf{p} = 1/r \partial_rU_r l$, the
valley-orbital splitting between the $p_+$ and $p_-$ states is
\begin{equation}
    \Delta \epsilon = -4\tau\frac{v^2}{\Delta^2}\int_0^\infty \frac{dU_r}{dr}\psi_r^2 dr , 
\end{equation}
where the p-wave radial function in the coordinate space is normalised as
\begin{equation}
    \int_0^\infty \psi_r^2 rdr = 1
\end{equation}
The wave function comes from the Fourier transform of $\psi_p$ from
Eq.~\eqref{eq:H_ex_p}, $\psi_{\bf p}=\psi_p e^{il\theta_p}.$
\begin{equation}
    \psi_r = \int_0^\infty \psi_p J_1(pr) \frac{pdp}{\sqrt{2\pi}}.
\end{equation}
Substituting in the Keldysh interaction, $v^2=\Delta/4\mu$ and
$\Delta = 1.9$ eV we find splitting $|\Delta \epsilon| = 3.8$ meV at $B=0$ T. Note that the sign of this splitting is dependent on valley index $\tau$. 
\bibliography{Exciton_and_TMD}
\end{document}